\documentstyle[12pt,aps,floats,prc,epsfig,tighten]{revtex}
\def\Journal#1#2#3#4{{ #1} {\bf #2}, #3 (#4) }
\def\NPA{{ Nucl. Phys.} {\bf A}}
\def\PRL{Phys. Rev. Lett.}
\def\PREV{ Phys. Rev.}
\def\PREP{ Phys. Rep.}
\def\PRC{{Phys. Rev.} C}
\def\PL {Phys. Lett.}
\begin{document}
\title{ Ground state of finite nuclei evaluated from  realistic interactions}
\author{Kh. Gad and  H. M\"uther}
\address{Institut f\"ur
Theoretische Physik, \\ Universit\"at T\"ubingen, D-72076 T\"ubingen, Germany}
\maketitle
 
\begin{abstract}
Ground state properties of finite nuclei ($^{16}$O and $^{40}$Ca) are evaluated
from realistic nucleon-nucleon interactions. The calculations are based on the
Brueckner-Hartree-Fock approximation. Special attention is paid to the role of
the energy spectrum for the particle states,  in particular for those close to
the Fermi energy. Additional binding energy is obtained from the inclusion of
the hole-hole scattering term within the framework of the Green function
approach. Results for the energy distribution of the single-particle strength
and the sensitivity to the nucleon-nucleon interaction are investigated. For
that purpose three modern nucleon-nucleon interactions are employed: the
Argonne V18, the charge dependent Bonn potential and a realistic
nucleon-nucleon interaction which is based on chiral perturbation theory and
which has recently been fitted by the Idaho group.
\end{abstract}

\section{Introduction}
One of the central aims of theoretical nuclear physics is the attempt to
determine the bulk properties of nuclear systems like its binding energy and
its density or radius from a realistic model of the nucleon-nucleon (NN)
interaction, i.e.~a NN interaction which yields an accurate fit of the NN
scattering data below the threshold of $\pi$-production. Quite some progress
has been made during the last ten years in the determination of such realistic
NN interaction models. A family of charge-dependent NN potentials has been
generated, which all fit the empirical scattering data with high
precision\cite{nijm1,argv18,cdb}. Nevertheless, using such NN interactions in 
nuclear structure calculations, one obtains results, which exhibit significant
differences\cite{nl1,nl2}. 

The NN interactions containing non-local terms like
the charge-dependent Bonn potential\cite{cdb}, CDBonn, or one of the Nijmegen
interaction model\cite{nijm1} tend to be ``softer'' than the purely local
interactions like the Argonne V18 potential\cite{argv18}. Here the expression
softer interaction is used to identify those interactions which induce weaker
two-nucleon correlations in the nuclear many-body wave function. This implies
that the total energy of nuclear matter calculated at a given density without
correlations, i.e.~using the Hartree-Fock approximation, yields less repulsive
results for a soft, non-local potential as compared to the energy calculated for
a stiffer, local NN interaction. As an example we mention that Hartree-Fock
calculations for nuclear matter at the empirical saturation density yield 4.6
MeV per nucleon using the  CDBonn interaction while 30.3 MeV are obtained
if the Argonne V18 potential is used\cite{nl2}. Including effects of NN
correlation as it is done e.g.~in a Brueckner-Hartree-Fock (BHF) calculation
leads to results which are attractive and much closer to each other. 

All these modern interactions, however, tend to predict too much binding energy
for nuclear matter and saturation densities which are too large. This is the
case for the BHF approach but is also observed if variational calculations are
performed\cite{akma1,brodb}  Softer non-local interaction yield larger binding
energies and saturation densities than local interactions\cite{nl2}. This
over-binding is often compensated by a three-nucleon force, which can be
adjusted in such a way that the empirical saturation point of nuclear matter is
reproduced\cite{grang,pudl,akma2,leje,zuo}. These three-nucleon forces can be
understood to simulate the relativistic effects as incorporated e.g.~in the
Dirac-BHF approach\cite{malf,brock,fuchs,schil}. These three-nucleon forces,
however, can also be interpreted to describe the effects of virtual excitations
of nucleons to the $\Delta$(3,3)\cite{fuji,delta} or $N^*$ Roper
resonance\cite{grang,shim}.

Phenomenological three-nucleon forces are also employed to describe the
properties of light nuclei with particle number up to A=8 using the Green
function Monte-Carlo method\cite{piper}. However, at first sight, the situation
in calculating bulk properties of finite nuclei seems to be different from the
corresponding situation in nuclear matter. While microscopic calculations for
nuclear matter, using the modern NN interactions, yield too much binding energy,
BHF calculations\cite{report} but also calculations using the coupled cluster
approach\cite{heisen} predict binding energies for finite nuclei, which are too
small as compared to the experimental data. 

It is the aim of the present work to investigate this situation a bit more in
detail. For that purpose we consider the BHF approximation paying special
attention to the particle-particle excitations at low energies and including the
effects of hole-hole scattering terms. We are investigating the
predictions originating from various models of realistic NN interactions. These
include local, like the Argonne V18\cite{argv18}, and non-local models, like
the CDBonn interaction\cite{cdb}, but also the realistic meson-exchange models
based on chiral perturbation theory, which have recently been developed by the
Idaho group\cite{idaho}.

After this introduction we will describe in section 2 a technique for the
calculation of bulk properties of nuclei which account in a consistent way 
for short-range and long-range correlations. Results of such calculation for the
nuclei $^{16}$O and $^{40}$Ca will be presented and discussed in section 3,
which is followed by a short summary in section 4. 

\section{Consistent treatment of long- and short-range correlations}

The BHF approximation is one of the most popular approaches to account for
effects of correlations beyond the mean-field approximation in calculating bulk
properties of infinite nuclear matter and finite nuclei from realistic NN
interactions. It is  characterized by a solution of the
Bethe-Goldstone equation, leading to the $G$-matrix ${\cal G_0}$
\begin{equation}
{\cal G_0(\omega )} = V + V \frac{Q_{0}}{ \omega - Q_{0} H_0
Q_{0}} {\cal G_0}\; ,
\label{eq:gmat0}
\end{equation}
and the self-consistent evaluation of the BHF single-particle energies 
\begin{equation}
\varepsilon_i = \left\langle i\left\vert \frac{p^2}{2m} \right\vert i
\right\rangle + \sum_{j<F} \left\langle ij\left\vert{\cal G_0}(\omega =
\varepsilon_i + \varepsilon_j)\right\vert ij \right\rangle\,.\label{eq:bhf}
\end{equation}
In these equations $V$ stands for the bare NN interaction, $Q_0$ represents the
Pauli operator, which restricts the propagator in the Bethe-Goldstone
Eq.(\ref{eq:gmat0}) to particle states with energies above the Fermi energy, and
$i$ and $j$ refer to hole-states, i.e.~the eigenstates of the BHF
single-particle hamiltonian with energies $\varepsilon_i$ and $\varepsilon_j$
below the Fermi energy. The self-consistent definition of the starting energy
$\omega$ in terms of the single-particle energies is determined by the
Bethe-Brandow-Petschek (BBP) theorem\cite{bbp}. Since, however, the BBP theorem
can only be used to justify the definition of the single-particle potential for
the hole states, the optimal definition of the single-particle energies for the
particle states, which enter the Bethe-Goldstone Eq.(\ref{eq:gmat0}) because
they define the eigenvalues of $H_0$, has been discussed over many years in the
literature. The conventional choice has been to ignore a single-particle
potential or  self-energy contributions for the particle states completely and
approximate $H_0$ by the kinetic energy only. This conventional choice is
supported by the coupled cluster or exponential S method\cite{kuem}, which
using the $S_2$ approximation essentially leads to the same approach,  This
conventional choice for the single-particle spectrum of nuclear matter is not
very appealing as it leads to a large gap at the Fermi surface.  

Mahaux and collaborators\cite{jeuk} argued that it would be more natural to
choose the propagator according to the Green function method, i.e.~define the
single-particle propagator with a single-particle energy which includes the real
part of the self-energy as a single-particle potential for particle and hole
states. This leads to a spectrum which is continuous at the Fermi momentum, which
provided the name ``continuous choice'' for this approach. This continuous choice
leads to an enhancement of correlation effects in the medium and tends to
predict larger binding energies for nuclear matter than the conventional choice.
Inclusion of three-hole-line contribution\cite{day,baldo} indicate that the
continuous choice seems to lead to a better convergence of the hole-line
expansion and is therefore preferable. In fact, recent studies in nuclear matter
show that the result is rather sensitive to details of the single-particle
spectrum around the Fermi energy\cite{fiasco,khaga1}.

In order to investigate these effects of the low-energy particle-particle 
excitations, which should correspond to long-range correlations, also in finite
nuclei, we follow the concept of a double partitioned Hilbert space, as it has
been used before for the study of infinite nuclear matter (see 
e.g.~\cite{kuo1,kuo2}) as well as finite nuclei (see e.g.~\cite{skouras,nili}).
The long range correlations are taken into account by means of the
Green function approach within a finite model space. This model
space shall be defined in terms of shell-model configurations
including oscillator single-particle states up certain shell. For our studies 
of $^{16}$O and $^{40}$Ca we have chosen to include configurations up to 
the pf shell and sdg shell, respectively. It turned out that the final results
are not very sensitive to this choice. This is due to the fact that the effect
of short-range correlations, i.e.~originating from configurations outside this
model space are not ignored but taken into account by means of an effective
interaction, i.e.~a $G$-matrix appropriate for this model space. This effective
interaction ${\cal G}$ is determined as the solution of the Bethe-Goldstone
equation
\begin{equation}
{\cal G}(\omega ) = V + V \frac{Q_{\hbox{mod}}}{ \omega - Q_{\hbox{mod}} H_0
Q_{\hbox{mod}}} {\cal G}\; .
\label{eq:gmat}
\end{equation} 
The Pauli operator $Q_{\hbox{mod}}$ is defined to exclude two-particle states
with one of the particles below the Fermi level of the nucleus considered or 
with both nucleons in states inside the model space. With this $G$-matrix we
solve the BHF Eq.(\ref{eq:bhf}) using the self-consistent definition of the
starting energy $\omega$ for hole states but also for the oscillator particle
states, which are inside the model space. We consider two different choices for 
the spectrum of the high-lying particle states outside the model space, defined
by $H_0$ in Eq.(\ref{eq:gmat}): The conventional choice, i.e.~$H_0$ is just the
operator of the kinetic energy for the interacting particles, and the continuous
choice, for which we add an attractive constant to the kinetic energy. This
constant is determined in such a way, that it corresponds to the mean value of
the potential energies of the low-lying  particle states inside the model space.

Note that this approximate BHF scheme only accounts for particle-particle
correlations outside the model space. We will call these correlations
short-range correlations  in the discussion below. A measure for  
these short-range correlations is given by the depletion coefficient
\begin{equation}
\kappa_i = -\sum_{j<F} \left\langle ij\left\vert\frac{\partial{\cal G}}{\partial
\omega} \right\vert ij \right\rangle\,\label{eq:kappa}
\end{equation}
or the corresponding occupation probability
\begin{equation}
\rho_i^{\mbox{sr}} = 1 -\kappa_i \,.\label{eq:rhosr}
\end{equation}
Using the single-particle energies we define an effective interaction for the
model space in terms of the oscillator matrix elements
\begin{equation}
\left\langle ij\left\vert{\cal V}\right\vert kl \right\rangle\ =
\frac{1}{2}\left[
\left\langle ij\left\vert{\cal G}(\omega =
\varepsilon_i + \varepsilon_j)\right\vert kl \right\rangle\ + 
\left\langle ij\left\vert{\cal G}(\omega =
\varepsilon_k + \varepsilon_l)\right\vert kl \right\rangle\ \right]\,.
\label{eq:veff}
\end{equation}

The effects of long-range correlations or correlations inside the model space
shall be evaluated by means of the Green function method. To determine the
correlated single-particle Green function for a nucleon with isospin $\tau$,
orbital angular momentum $l$ and total angular momentum $j$  one has to solve a
Dyson equation, which in the case of a finite system with spherical symmetry and
within a discretize model space takes the form
\begin{equation}
g_{\tau lj}(n,n';\omega ) = g^{(BHF)}_{\tau lj}(n,n';\omega )
+ \sum_{n'',n'''} g^{(BHF)}_{\tau lj}(n,n'';\omega ) \Delta\Sigma_{\tau lj}
(n'',n''';\omega )  g_{\tau lj}(n''',n';\omega ) \,,
\label{dyso}
\end{equation}
where $g^{(BHF)}_{\tau lj}(n,n'';\omega )$ refers to the BHF propagator, which
we assume to be diagonal in the radial quantum numbers $n$, $n'$ 
\begin{equation}
g^{(BHF)}_{\tau lj}(n,n';\omega ) = \delta_{n,n'}\frac{1}{\omega -
\varepsilon_{n\tau l j} \pm i \eta}\,.\label{eq:gbhf}
\end{equation}
The correction to the BHF
self-energy in terms of two particle one hole ($2p1h$) and two hole one particle
($2h1p$) configurations inside the model space
\begin{equation}
\Delta\Sigma_{\tau lj}(n,n',\omega) = \Sigma_{\tau lj}^{(2p1h)}(n,n',\omega)
+ \Sigma_{lj}^{(2h1p)}(n,n',\omega) \, ,
\label{self}
\end{equation}
is defined in terms of the effective interaction ${\cal V}$ of
Eq.(\ref{eq:veff}). As an example we consider the $2p1h$ contribution of second
order in ${\cal V}$
\begin{equation}
\Sigma_{\tau lj}^{(2p1h)}(n,n',\omega )  =
 \frac{1}{2} \enspace
\sum_{h<F} \enspace \sum_{p_{1},p_{2}>F} \enspace \frac
{< nh \vert {\cal V}\vert p_{1} p_{2}>
<  p_{1} p_{2} \vert {\cal V}\vert n' h>}
{\omega - (\varepsilon_{p1}+\varepsilon_{p2}-\varepsilon_h) + i\eta }\,.
\label{part}
\end{equation}
In order to evaluate the total energy of the system and expectation values of
one-body operators one has to rewrite the single-particle Green function in the
Lehmann representation
\begin{equation} 
g_{\tau lj}(n,n';\omega ) = \sum_\alpha \frac{S_{\tau
lj}^p(n,n',\omega_{\alpha\tau  lj})}{\omega - \omega_{\alpha\tau lj}+i\eta} +
\sum_\beta \frac{S_{\tau lj}^h(n,n', \omega_{\beta\tau lj})}{\omega -
\omega_{\beta\tau lj}-i\eta}\,.\label{eq:lehm}
\end{equation}
The single-particle density matrix is then defined in terms of the hole-spectral
function
\begin{equation}
\rho_{\tau lj}(n,n') = \sum_\beta S_{\tau lj}^h(n,n',\omega_{\beta\tau
lj})\label{eq:rho}
\end{equation} 
and the total energy can be evaluated from 
\begin{equation}
E = \sum_{\tau l j\beta n,n'} \frac{(2j+1)}{2} S_{\tau
lj}^h(n,n',\omega_{\beta\tau  lj})\left\langle  n\left\vert \frac{p^2}{2m} +
\omega_{\beta\tau lj}\right\vert n'\right\rangle \,.\label{eq:koltun}
\end{equation}
In order to obtain the spectral function and the positions of the poles in the
single-particle Green function, $\omega_{\beta\tau lj}$, we reformulate the
Dyson equation into an eigenvalue problem \cite{report,skouras}
\begin{equation}
\pmatrix{\varepsilon_1& & 0& a_{11} & \ldots & a_{1P} & A_{11} &
\ldots & A_{1Q} \cr
 & \ddots&  & \vdots & & \vdots &\vdots & & \vdots \cr
0 & &\varepsilon_N & a_{N1} & \ldots & a_{NP} & A_{N1} &
\ldots & A_{NQ} \cr
a_{11} & \ldots & a_{N1} & e_1 & & & 0 & & \cr \vdots & &\vdots & & \ddots
& & &  & \cr
a_{1P} & \ldots & a_{NP} & 0 & & e_P & & & 0 \cr
A_{11} & \ldots & A_{N1} &  & & &  E_1 & & \cr
\vdots & &\vdots& & & & & \ddots & \cr
A_{1Q} & \ldots &A_{NQ}& 0 & \ldots & 0 & & \ldots & E_Q \cr }
\pmatrix{ X_{0,1}^\alpha \cr \vdots \cr X_{0,N}^\alpha \cr X_1^\alpha \cr \vdots
\cr X_P^\alpha \cr Y_1^\alpha \cr \vdots \cr Y_Q^\alpha
\cr }
= \omega_\alpha
\pmatrix{ X_{0,1}^\alpha \cr \vdots \cr X_{0,N}^\alpha\cr X_1^\alpha \cr 
\vdots \cr
X_P^\alpha \cr Y_1^\alpha \cr \vdots \cr Y_Q^\alpha
\cr } \; ,
\label{eigen}
\end{equation}
where for simplicity we have dropped the corresponding conserved
quantum numbers for isospin, parity and angular momentum ($\tau lj$).
The matrix to
be diagonalized contains the BHF hamiltonian defined in terms of the $N$ BHF
single-particle energies of the symmetry assumed within the model space, the
 coupling to the $P$ different $2p1h$
configurations and $Q$ $2h1p$ states. These couplings are expressed in terms of
the matrix elements
\begin{equation}
a_{mi}   = < m h \vert{\cal V} \vert p_{1} p_{2}>\qquad \mbox{and}\qquad
A_{mj}   = < m p \vert{\cal V} \vert h_{1} h_{2}>\,.
\label{mael}
\end{equation}
As long as we are still ignoring any residual interaction between the various
$2p1h$ and $2h1p$ configurations the corresponding parts of the matrix in
(\ref{eigen}) are diagonal with elements in terms of single-particle energies
and denoted by $e_i$ ($E_j$)  for $2p1h$ ($2h1p$) configurations, respectively.
One can easily improve this approach and incorporate the effects of residual
interactions between the $2p1h$ configurations or $2h1p$ configurations. One
simply has to modify the corresponding parts of the matrix in Eq.(\ref{eigen})
and replace where ${\cal H}_{2p1h}$ and ${\cal H}_{2h1p}$ contain
the residual interactions in the $2p1h$ and $2h1p$
subspaces 
\begin{equation}
\pmatrix{e_{1} & \ldots & 0 \cr \vdots & \ddots &\cr 0 &\ldots &e_{P}\cr}
\Longrightarrow {\cal H}_{2p1h} \; ,\qquad\mbox{and}\qquad
\pmatrix{E_{1} & \ldots & 0 \cr \vdots & \ddots &\cr 0 &\ldots &E_{Q}\cr}
\Longrightarrow {\cal H}_{2h1p} \; .
\label{restw}
\end{equation}
For the calculations which we are going to discuss here, we only consider the
matrix elements for the particle-particle interaction in ${\cal H}_{2p1h}$ and
the hole-hole interaction for ${\cal H}_{2h1p}$. This implies that the
corresponding particle-particle and hole-hole ladder diagrams are taken into
account. A more complete treatment of the residual interaction requires the
treatment of three-body terms and has recently been discussed e.g.~by Barbieri
et al.\cite{carlob}  

The eigenvalues $\omega_\alpha$ of Eq.(\ref{eigen}) correspond to the poles of
the single-particle Green function in Eq.(\ref{eq:lehm}) and the spectral
function is given in terms of the components of the eigenvectors by
\begin{equation}
S^h(n,n';\omega_\alpha) = X_{0,n}^\alpha X_{0,n'}^\alpha \label{eq:speccal}
\end{equation} 
for eigenvalues $\omega_\alpha$ below the Fermi energy $E_F$, while a
corresponding equation holds for the spectral function $S^p$ for energies above
$E_F$.

\section{Results and discussion}

As a first example we would like to consider some results displayed in table
\ref{tab1} evaluated for the nucleus $^{16}$O using the CDBonn
potential\cite{cdb} supplemented by the Coulomb interaction between protons. 
For the results displayed in this table the
single-particle wave functions have been constrained to wave functions of the
harmonic oscillator defined in terms of an harmonic oscillator constant 
\begin{equation}
b = \frac{1}{\sqrt{2}\alpha}\,, \label{eq:balpha}
\end{equation}
with $\alpha$ = 0.4 fm$^{-1}$. This corresponds to an oscillator frequency of
\begin{equation}
\hbar\omega = \frac{(\hbar c)^2}{mc^2b^2}\label{eq:hbomeg}
\end{equation}
of 13.27 MeV and leads to a radius for $^{16}$O, assuming simple shell model
occupancies, of 2.65 fm, which is close to the empirical value. The first two
columns of table \ref{tab1} refer to BHF calculation assuming the conventional
and a continuous choice for the spectrum of the particle states in the
Bethe-Goldstone equation (\ref{eq:gmat0}). It should be recalled that we define
the continuous spectrum in terms of the kinetic energy shifted by a constant
such that the single-particle energies for low-lying particle  states is
identical to the mean value of the corresponding BHF single-particle energies
calculated according to (\ref{eq:bhf}). Beside the single-particle energies
also the occupation probabilities $\rho_i^{\mbox{sr}}$ calculated according to
Eq. (\ref{eq:rhosr}) are listed. Comparing these two columns one can see that
the use of a continuous single-particle spectrum leads to  more attractive
single-particle energies and a binding energy, which is enhanced by about 2 MeV
per nucleon. This enhancement of the attraction is accompanied by lower
occupation probabilities $\rho_i^{\mbox{sr}}$, which implies larger values for
the corresponding depletion coefficients. This demonstrates that the lowering
of the particle-state spectrum, going from the conventional to the continuous
choice, leads to a substantial enhancement of correlations. 

The third and fourth column of table \ref{tab1}, labeled BHF0 and BHF(mod), 
refer to the model space approach as introduced in the preceding section. The
BHF0 approach identifies the BHF calculation using the 
G-matrix defined in (\ref{eq:gmat}) with a continuous choice for the particle
state spectrum. This means that due to the Pauli operator $Q_{\mbox{mod}}$ the
contribution of low-lying particle states (for the present example those in the
1s0d and 1p0f shell) to the G-matrix are suppressed. This leads to larger
occupation probabilities (compare BHF0 with BHF (cont)) and less
attractive single-particle energies. Also the binding energy per nucleon is
reduced by about 1 MeV.

The contribution of the low-lying particle-particle configurations are again
included in the BHF(mod) approach by considering the corresponding (2p1h)
component in the definition of $\Delta \Sigma$ in Eq.(\ref{self}) and solving
Eq.(\ref{eigen}) ignoring the coupling to the (2h1p) configurations but
allowing for the residual interaction between the particle states according to
Eq.(\ref{restw}). In this way the effects of the low-lying particle-particle
configurations are taken into account allowing for individual single-particle
energies $\varepsilon_{nlj}$ for all sub-shells with quantum numbers $nlj$ and
not just a replacement of these single-particle energies with the kinetic
energy shifted by a global constant as it was done in the BHF (cont) approach
discussed above. This more sophisticated treatment of the low-lying particle
spectrum leads to some additional attraction in the single-particle energies
ranging from 0.9 MeV in the case of the $s_{1/2}$ shell over 1.6 MeV
($p_{3/2}$) to 1.8 MeV in the case of the $p_{1/2}$ shell. The effect is
obviously larger for states closer to the Fermi energy as these states are more
sensitive to the details of the long-range correlations. The same feature can
also be observed in the occupation probabilities $\rho_{lj}$. It should be
pointed out that the results listed for the BHF (mod) approach accounts for
depletion due to the excitation of particle state configurations inside the
model space by means of Eq.(\ref{eq:rho}) while the depletion due to
short-range correlations leading to excitations outside the model space are
accounted for by means of Eq.(\ref{eq:rhosr}). The more specific treatment of
the low-lying particle-particle configuration reduces the spin-orbit splitting
for the protons in the $p$-shell from 3.4 MeV in the case of BHF(cont) to 3.1
MeV in the BHF(mod) approach, both values being much smaller than the
experimental one (6.3 MeV).

\begin{figure}[ht]
\begin{center}
\epsfig{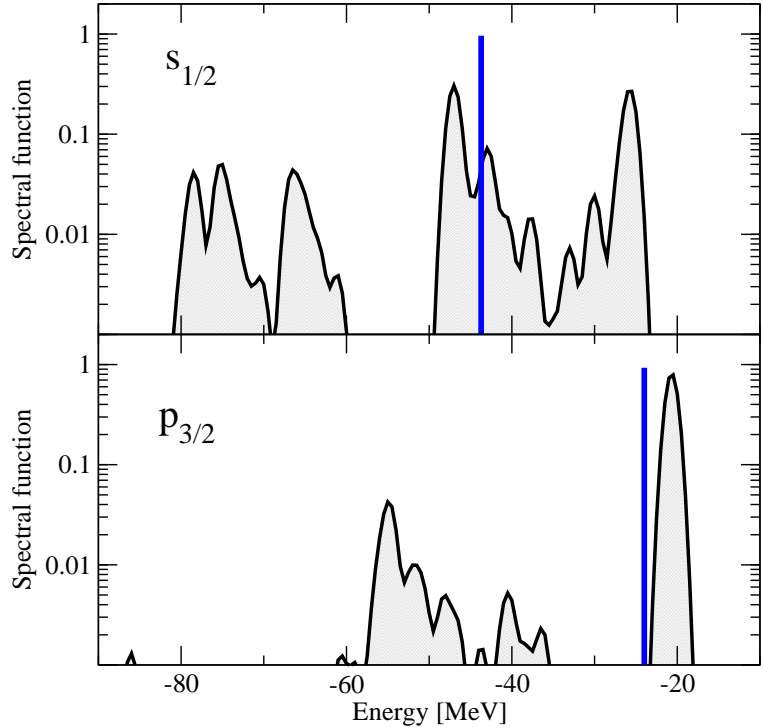}
\end{center}
\caption{The spectral function for proton hole strength in the 
$s_{1/2}$ (upper panel) and $p_{3/2}$ channel. The results are calculated for
$^{16}$O using the CD Bonn interaction. The distributions of the complete Green
function approach 
are obtained by folding the discrete distribution with Gaussian functions 
assuming a width of 1 MeV. Also given are the positions of the single-particle 
states evaluated in the BHF(mod)  approximation (straight lines).   
\label{fig1}}
\end{figure}

For all approaches discussed so far, the single-particle strength for the hole
states is just concentrated in one quasiparticle state. The corresponding Green
functions exhibit only one pole below the Fermi energy. A distribution of this
hole-strength is obtained if we also account for the 2h1p contribution in the
definition of the self-energy in Eq.(\ref{self}). This distribution is defined
in terms of the eigenvalues $\omega_\alpha$ in Eq.(\ref{eigen}) and the
strength defined in Eq.(\ref{eq:speccal}). Results for the spectral
distribution are displayed in Fig.~\ref{fig1}. In the upper panel of this
figure, referring to the removal of protons  with $s_{1/2}$ quantum numbers,
the strength is widely distributed.  In the column labeled ``Green'' of table
\ref{tab1} we give the mean value of this spectral distribution defined by
\begin{equation}
\varepsilon_{\tau lj} = \frac{\sum_{\beta n} \omega_{\beta\tau lj}
S_{\tau lj}^h(n,n,\omega_{\beta\tau lj})}{\sum_n \rho_{\tau lj}(n,n)}
\label{eq:emean}
\end{equation} 
but also the energy of the quasiparticle state (in brackets), which is defined 
by that eigenvalue $\omega_{\beta\tau lj}$, which carries the largest strength.
Inspecting Fig. \ref{fig1} and the corresponding numbers in table \ref{tab1}, 
one finds that the quasiparticle state for the $s_{1/2}$ state carries only
rather little strength. 

The quasiparticle contribution is much more important for the $p_{3/2}$ and 
$p_{1/2}$ states. For these more weakly bound states the quasiparticle states
are also those which are closest to the Fermi energy, which means that they
correspond to the removal of a nucleon leading to the ground state or lowest
excited state of the daughter nucleus of given parity ($l$) and angular momentum
($j$). Therefore one should consider those quasiparticle energies in comparing
with experimental removal energies for these states (values presented in the
last column of table \ref{tab1}). The spin-orbit splitting resulting from these
quasiparticle energies is slightly larger than the one derived from BHF(mod)
approximation but
still too small as compared with the experimental value. Here it should be
recalled that a substantial enhancement of the spin-orbit splitting is obtained
if the relativistic features of the Dirac Brueckner Hartree Fock approximation
are taken into account\cite{spinorbit}.

Comparing the results of the Green function approach with those obtained in the
BHF(mod) approximation, one finds that the hole-hole terms which are included in
the Green function approach tend to reduce the binding energy of the 
quasiparticle state but lead to more attractive mean values of the spectral 
distribution for the hole states. Since this mean value enters the evaluation of
the total binding energy (see Eq.(\ref{eq:koltun})) we also get a slightly 
larger binding energy in the Green function approach as compared to the 
BHF(mod) approximation. It is worth mentioning that the Green function approach
with inclusion of 2h1p terms in the self-energy also provides non-vanishing
occupation probabilities for states which are unoccupied in the simple shell
model. Also these occupancies contribute to the total binding energy.

At this stage it is useful to make a first comparison with the situation in
corresponding calculations of nuclear matter. It has been observed also for
nuclear matter that the results of BHF calculations are rather sensitive to the
choice of the spectrum for the particle states, in particular for those with
momenta close to the Fermi momentum\cite{fiasco,khaga1}. Using the precise
single-particle spectrum rather than a quadratic parametrisation leads to an
increase of the binding energy in nuclear matter around saturation density of 
about 1.5 MeV per nucleon\cite{khaga1}. This is even more than the gain in
binding energy from the BHF(cont) to the BHF(mod) approach displayed in table
\ref{tab1}. Also in nuclear matter one observes that the inclusion of the 2h1p
terms in the self-energy lead to less attractive quasiparticle energies. The
spread of the single-particle strength to lower energies, however, leads to
additional binding energy. This feature, less attractive quasiparticle energies
but more binding energy per nucleon, helps to fulfill the Hugenholtz - Van
Hove  theorem\cite{hvh,alfredo} for nuclear matter. The gain in binding energy
due to the 2p1h components in the self-energy is about 0.5 MeV per nucleon in
nuclear matter at saturation density\cite{khaga1}. Also this is a result rather
similar to the difference between the BHF(mod) approach and the result of the
complete Green function listed in table \ref{tab1}.

For a first comparison of results originating from different NN interactions,
we list in table \ref{tab2} results of the Green function calculation for
$^{16}$O using the charge-dependent Bonn interaction\cite{cdb} (CD Bonn), the
version A of the Idaho interaction\cite{idaho} (Idaho A), which is based on
features of chiral perturbation theory, and the Argonne V18
interaction\cite{argv18}. All interaction models yield quasiparticle energies
which are more attractive than the experimental removal energies but a total
binding energy which is slightly smaller than the empirical energy of 7.98 MeV
per nucleon. 

The prediction for the spectral strength of the $p_{3/2}$ and $p_{1/2}$
quasiparticle states range between 0.71 and 0.75. These values are 
larger than spectroscopic factors deduced from nucleon knock-out experiments,
$(e,e')p$, which are around 0.6\cite{leusch}.
The local NN interaction Argonne V18 is stiffer than the non-local
meson-exchange interaction models CD Bonn and Idaho A (see also the comparison
in nuclear matter \cite{khaga1}). This means that it predicts stronger
correlations, as indicated by the smaller spectroscopic factors, and smaller
binding energy.

\begin{figure}[ht]
\begin{center}
\epsfig{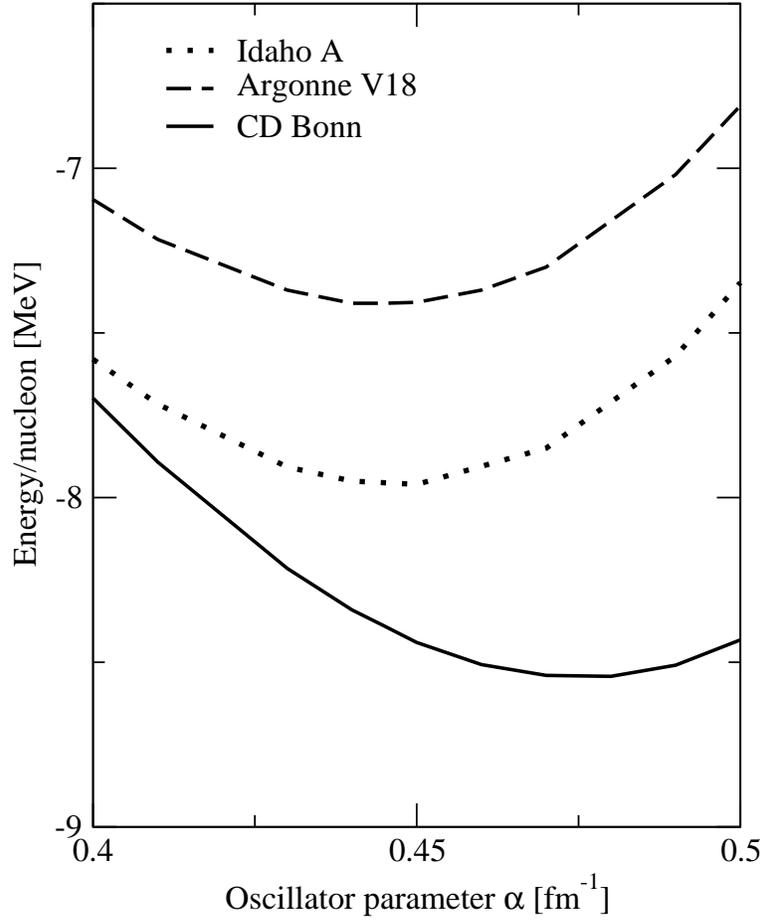}
\end{center}
\caption{Binding energy per nucleon calculated in the Green function approach
as a function of the oscillator parameter $\alpha$ (see Eq.(
\protect{\ref{eq:balpha}}), which is used to define the basis of the model
space). Results are displayed for various NN interactions \label{fig2}}   
\end{figure}

Altogether one may argue that the bulk properties of $^{16}$O are very well
reproduced. All interactions reproduce the total energy within 1 MeV per nucleon
and predict a radius for the proton distribution of 2.72 fm, which is in very
good agreement with the experiment. Here, however, one must keep in mind that
the radius is to a large extent determined by the choice of the oscillator
parameter for the model space. The choice of $\alpha =0.4$ has been made to
obtain a radius of 2.65 fm for the uncorrelated shell-model wave function.
Correlations within the model space lead to small enhancements, only. In order
to test the sensitivity of the calculation on the oscillator parameter, we have
performed calculations for various values of $\alpha$. Results for the binding
energy of $^{16}$O obtained in the Green function approach are displayed in Fig.
\ref{fig2} as a function of the oscillator parameter $\alpha$. These figures
show that an energy minimum is obtained for $\alpha$ around 0.475 fm$^{-1}$ 
(0.45 fm$^{-1}$, 0.44 fm$^{-1}$) using the CD Bonn, the Idaho A and the Argonne
V18 interaction, respectively.  The corresponding radii of the proton
distribution are 2.29 fm, 2.41 fm and 2.48 fm, all of them to small as compared
to the experimental value. Therefore the situation for calculating bulk
properties of $^{16}$O is similar to the attempts of evaluating the saturation
point of nuclear matter. The calculation tend to predict a radius which is to
small or a density which is too large. It is worth noting, however, that the
disagreement for nuclear matter is larger. 

As a second example we also consider the nucleus $^{40}$Ca. In this case we
consider a model space, which is defined in terms of harmonic oscillator wave
functions including all shells up to the 2s1d0g shell. For a first comparison we
fix the oscillator constant to $\alpha$ = 0.35 fm$^{-1}$. This corresponds to
simple shell model prediction for the radius of 3.5 fm, in reasonable agreement
with the empirical value of 3.48 fm. 

\begin{figure}[ht]
\begin{center}
\epsfig{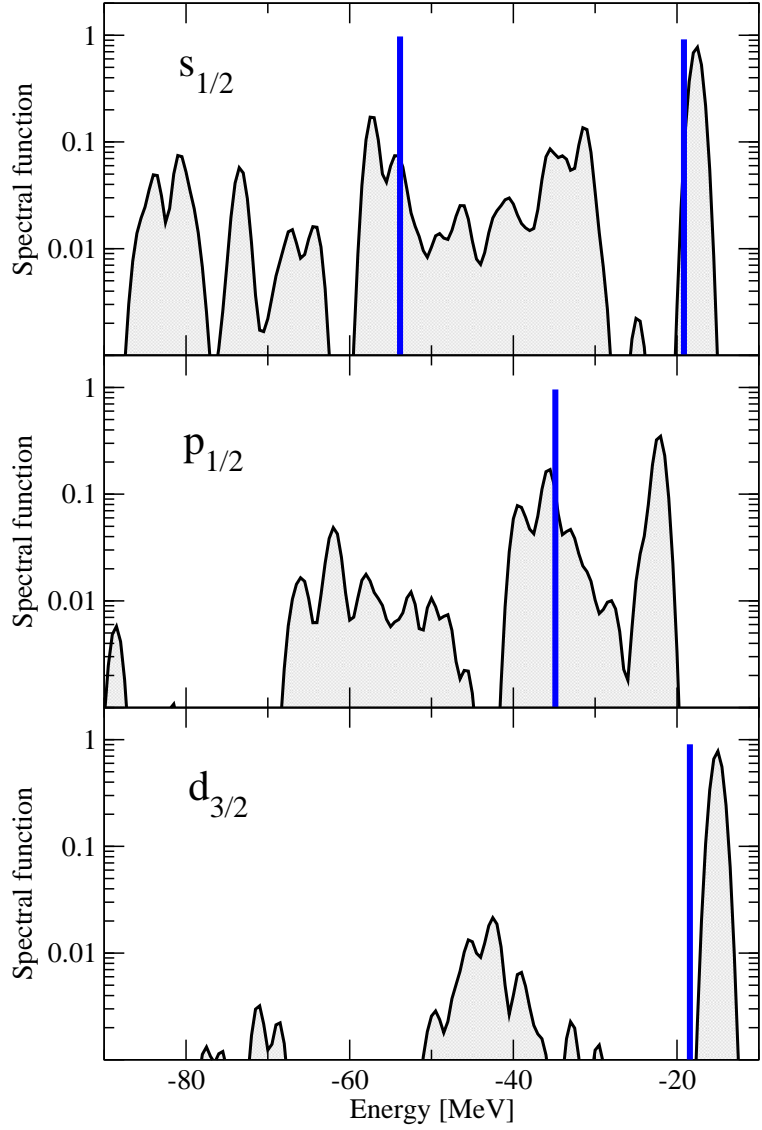}
\end{center}
\caption{The spectral function for proton hole strength in the 
$s_{1/2}$ (upper panel) and $p_{1/2}$ (middle) and $d_{3/2}$ channel. 
The results are calculated for
$^{40}$Ca using the CD Bonn interaction. Further details see Fig.
\protect\ref{fig1}.   
\label{fig3}}
\end{figure}

Examples for the spectral distribution are displayed in Fig.~\ref{fig3}
(assuming the CD Bonn interaction) and numerical results obtained in the Green
function approach are listed in table \ref{tab3}. Since we have two
quasiparticle states for the $s_{1/2}$ channel in the BHF approximation (see
upper panel of Fig.~\ref{fig3}), we also list two energies and occupation
probabilities for this partial wave in table \ref{tab3}. Also for this nucleus
one obtains a broad distribution of the spectral strength. A well defined
quasiparticle peak only shows up for the states in the 1s0d shell.

Also in this case, the predictions for the removal energy (absolute value of the
quasiparticle energy for $d_{3/2}$, are larger than the experimental value of
8.3 MeV. Note, however, that the inclusion of 2h1p contributions in the Green
function approach reduces the discrepancy by more than 3 MeV as compared to the
BHF(mod) approximation. On the other hand, the calculated binding energies are
in good agreement with the experimental value of 8.55 MeV per nucleon. So again,
the calculation of bulk properties of $^{40}$Ca is rather successful, if one
fixes the radius with the appropriate choice of the oscillator constant
$\alpha$, which defines the basis of the model space. 

\begin{figure}[ht]
\begin{center}
\epsfig{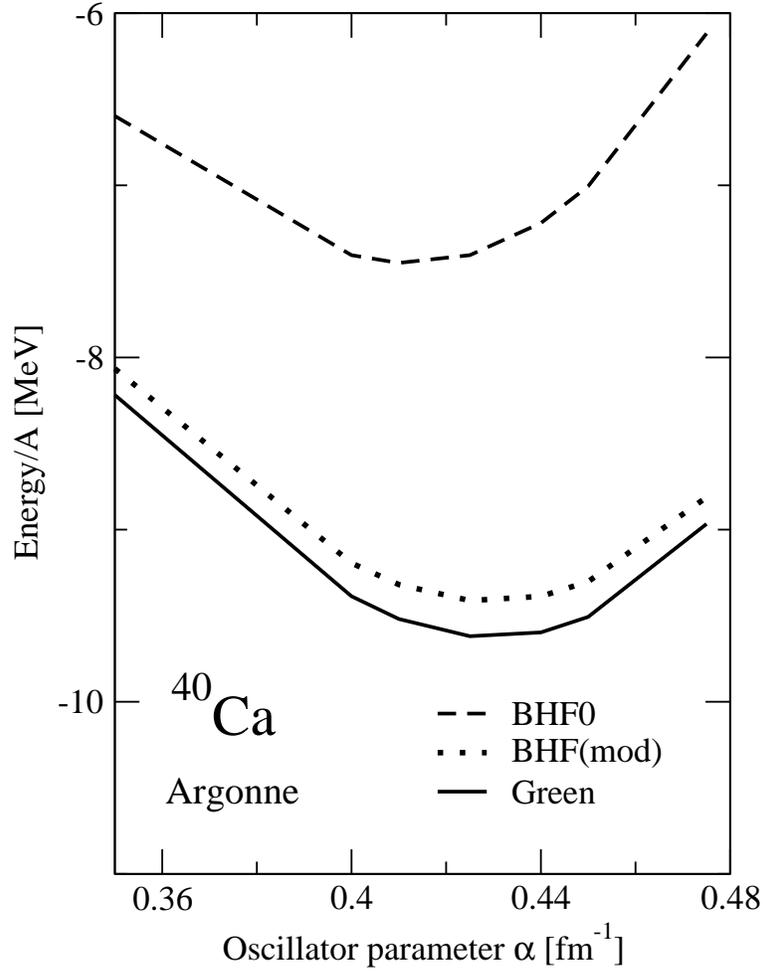}
\end{center}
\caption{Binding energy per nucleon for $^{40}$Ca calculated in the Green
function approach as a function of the oscillator parameter $\alpha$. Results
are displayed for various approximation schemes (for nomenclature see table
\protect{\ref{tab1}}) \label{fig4}}   
\end{figure}

\begin{figure}[ht]
\begin{center}
\epsfig{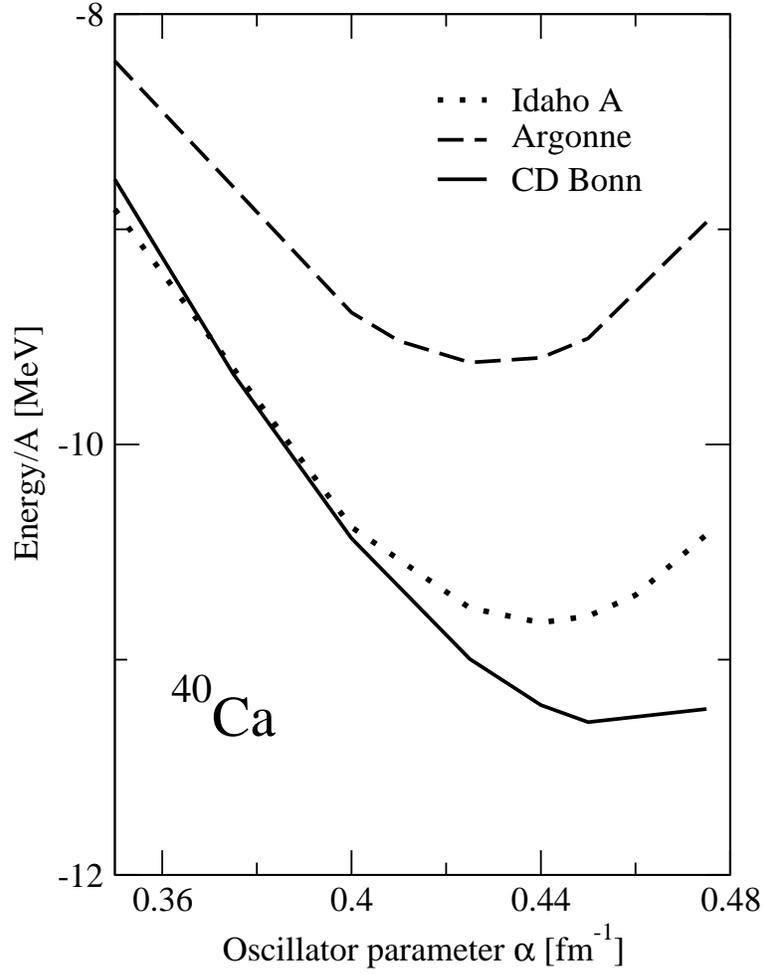}
\end{center}
\caption{Binding energy per nucleon calculated in the Green function approach
as a function of the oscillator parameter $\alpha$.
Results are displayed for various NN interactions   
\label{fig5}}
\end{figure}

If one releases this constraint and considers various values for $\alpha$ one
obtains results as displayed in Figs.~\ref{fig4} and \ref{fig5}. Also in this
case we observe that the minima occur for oscillator parameters which are larger
than $\alpha$ = 0.35 fm$^{-1}$, which means that the corresponding radii are
smaller than the empirical one. This is true for the various approximations (see
Fig.~\ref{fig4}) but also for the various interactions (see Fig.~\ref{fig5}).  

The comparison of the various approximation schemes in Fig.~\ref{fig4} confirms
the conclusions, which have been given above for the case of $^{16}$O. In the
case of $^{40}$Ca, however, all interactions predict too much binding energy
and a radius, which in the case of CD Bonn interaction is about 25 percent
smaller  than the experimental value. This corresponds to an average density,
which is too large by about a factor of two, a situation which is approaching
the situation of nuclear matter. Also in this case, the stiffer interaction
Argonne V18 yields less energy and smaller densities than the softer
interactions (Idaho A and CD Bonn).

\section{Conclusions} 
Starting from the Brueckner-Hartree-Fock (BHF) approach various approximation
schemes have been investigated to derive bulk properties of finite nuclei from
realistic NN interactions. It is observed that the results of BHF calculations,
for finite nuclei as well as infinite nuclear matter\cite{khaga1},
are rather sensitive to the spectrum of particle states, in particular those 
with energies close to the Fermi energy. Therefore a technique has been applied,
to separate a detailed description of the low-energy excitations corresponding
to long range correlations from the treatment of short-range correlations. While
the effects of short-range correlations are taken into account by means of the
G-matrix approximation of the BHF scheme, the long-range correlations are
considered within the framework of the self-consistent evaluation of
single-particle Green function. This approach includes the effects of
particle-particle correlations but also corresponding hole-hole scattering
terms. 

If the basis of the model space is constrained to obtain the empirical value for
the radius, one is able to reproduce the binding energy of nuclei (we consider
$^{16}$O and $^{40}$Ca) within 0.5 MeV per nucleon, which is very satisfactory
for a calculation based on realistic NN interaction. This success is due to the
attraction which is obtained from the careful treatment of the low-lying
particle-particle excitations but also due to the inclusion of two-hole
one-particle configurations in the definition of the self-energy for the
single-particle propagator. These terms yield a distribution of the 
single-particle strength and a shift of the quasiparticle energy. The repulsive
shift of the quasiparticle energy for states close to the Fermi level improves
the agreement with the experimental removal energies. The distribution of the
strength, on the other hand, allows a gain in binding energy although the
quasiparticle energies are less attractive. This helps to improve the
fulfillment of the Hugenholtz - Van Hove theorem. The calculated spectroscopic
factors for the quasiparticle states (around 0.7) are slightly above the
empirical values (around 0.6) derived from $(e,e')p$ experiments.

If the constraining condition for the radius is released, the minima for the
calculated energies lead to radii which are too small and densities which are
too large. The calculated binding energies tend to be larger than the empirical
values. This is more pronounced for the nucleus $^{40}$Ca than for $^{16}$O. 
This deficiency could be cured by considering the repulsive three nucleon
forces\cite{grang,pudl,akma2,leje,zuo}, which are adjusted to describe the
saturation point of infinite nuclear matter also for the calculation of finite
nuclei.

\vfil\eject
\begin{table}[ht]
\begin{center}
\begin{tabular}{c|rrrrr|r}
& BHF (conv) & BHF (cont)& BHF0 & BHF (mod) & Green & Exp\\
\hline
$\varepsilon_{s1/2}$ & -38.19 & -42.78 & -40.74 & -43.72 & -44.00 (-47.01) 
& -40 $\pm$ 8 \\
$\varepsilon_{p_3/2}$ & -18.14 & -22.40 & -20.40 & -23.99 & -24.29 (-20.68) 
& -18.45 \\
$\varepsilon_{p_1/2}$ & -14.50 & -19.02 & -16.86 & -20.86 & -21.26 (-17.44) 
& -12.13 \\
\hline
$\rho_{s1/2}$ & 0.928 & 0.854 & 0.904 & 0.871 & 0.867 (0.276) & \\
$\rho_{p3/2}$ & 0.926 & 0.855 & 0.898 & 0.829 & 0.823 (0.730) & \\
$\rho_{p1/2}$ & 0.916 & 0.857 & 0.900 & 0.818 & 0.802 (0.725) & \\
\hline
E/A & -4.61 & -6.67 & -5.66 & -7.57 & -7.78 & -7.98 \\
\end{tabular}
\caption{\label{tab1} Single-particle energies ($\varepsilon_i$) and occupation
probabilities ($\rho_i$) for protons in $^{16}$O and the total energy per
nucleon (E/A) as calculated from various approximations (see discussion in the
text) are compared to experimental data. The calculations have been performed
using the CD Bonn interaction and considering a model space defined in terms of
oscillator functions with parameter $\alpha =0.4$. The numbers in brackets refer
to energy and occupation probability of the dominant quasiparticle contribution
in the Green function approach. All energies are given in
MeV.} 
\end{center}
\end{table}
\begin{table}[ht]
\begin{center}
\begin{tabular}{c|rrr}
& CD Bonn & Idaho A & Arg. V18 \\
\hline
$\varepsilon_{s1/2}$ & -44.00 (-47.01) & -44.00 (-47.43) & -42.38 (-45.63) 
\\
$\varepsilon_{p_3/2}$ & -24.29 (-20.68) & -24.00 (-20.64) & -23.12 (-19.77) \\
$\varepsilon_{p_1/2}$ & -21.26 (-17.44) & -20.87 (-17.28) & -20.26 (-16.62) \\
\hline
$\rho_{s1/2}$ & 0.867 (0.276) & 0.873 (0.284) & 0.830 (0.265)\\
$\rho_{p3/2}$ & 0.823 (0.730) & 0.833 (0.747) & 0.806 (0.723)\\
$\rho_{p1/2}$ & 0.802 (0.725) & 0.821 (0.741) & 0.794 (0.715)\\
\hline
E/A & -7.78 & -7.65 & -7.15  \\
\end{tabular}
\caption{\label{tab2} Results for single-particle energies, occupation
probabilities and total energy per nucleon of $^{16}$O calculated from the Green
function approach using the CD Bonn, the Idaho A and the Argonne V18 interaction
model. Further details see table \protect\ref{tab1}.} 
\end{center}
\end{table}
\begin{table}[ht]
\begin{center}
\begin{tabular}{c|rrr}
& CD Bonn & Idaho A & Arg. V18 \\
\hline
$\varepsilon_{s1/2}$ & -38.30 (-57.26) & -38.81 (-58.44)& -37.92 (-55.53)\\
& (-17.63) & (-17.94) & (-16.84) \\
$\varepsilon_{p_3/2}$ & -37.59 (-23.22) & -38.06 (-23.76) & -36.37 (-22.52)\\
$\varepsilon_{p_1/2}$ & -35.17 (-22.16) & -35.60 (-22.59) & -34.12 (-21.42)\\
$\varepsilon_{d_5/2}$ & -22.27 (-18.35) & -22.42 (-18.70) & -21.33 (-17.61)\\
$\varepsilon_{d_3/2}$ & -18.88 (-15.08) & -18.96 (-15.34) & -18.13 (-14.46)\\
\hline
$\rho_{s1/2}$ & 1.658 (0.160) & 1.674 (0.177)& 1.592 (0.142)\\
& (0.697) & (0.715) & (0.682)  \\
$\rho_{p3/2}$ & 0.834 (0.123) & 0.844 (0.141)& 0.800 (0.151)\\
$\rho_{p1/2}$ & 0.826 (0.301) & 0.838 (0.309)& 0.794 (0.292)\\
$\rho_{d5/2}$ & 0.813 (0.691) & 0.824 (0.709)& 0.795 (0.679)\\
$\rho_{d3/2}$ & 0.796 (0.701) & 0.808 (0.717)& 0.780 (0.688)\\
\hline
E/A & -8.77 & -8.91 & -8.22 \\
\end{tabular}
\caption{\label{tab3} Results for single-particle energies, occupation
probabilities and total energy per nucleon of $^{40}$Ca calculated from the Green
function approach using the CD Bonn, the Idaho A and the Argonne V18 interaction
model. Further details see table \protect\ref{tab1}.} 
\end{center}
\end{table}

\begin{thebibliography}{99}
\bibitem{nijm1} V.G.J.\ Stoks, R.A.M.\ Klomp, C.P.F.\ Terheggen,
and J.J.\ de Swart, \Journal{\PRC}{49} {2950} {1994}.
\bibitem{argv18} R.B.\ Wiringa, V.G.J.\ Stoks, and R.\ Schiavilla,
\Journal{\PRC} {51} {38} {1995}.
\bibitem{cdb} R.\ Machleidt, F.\ Sammarruca, and Y.\ Song, \Journal{\PRC}{53}
{R1483} {1996}.
\bibitem{nl1} A. Polls, H. M\"uther, R. Machleidt, and M. Hjorth-Jensen,
\Journal{\PL}{B 432}{1}{1998}.
\bibitem{nl2} H. M\"uther and A. Polls, \Journal{\PRC} {61} {014304} {2000}.
\bibitem{akma1} A. Akmal and V.R. Pandharipande, \Journal{\PRC} {56} {2261}
{1997}.
\bibitem{brodb} G.H. Bordbar and M. Modarres, \Journal{\PRC} {57} {714}
{1998}.
\bibitem{grang} P. Grange, A. Lejeune, M. Martzolff, and J.F. Mathiot, 
\Journal{\PRC} {40} {1040}{1989}.
\bibitem{pudl} B.S. Pudliner, V.R. Pandharipande, J. Carlson, and R.B. Wiringa,
\Journal{\PRL} {74}{4396}{1995}.
\bibitem{akma2} A. Akmal, V.R. Pandharipande, and D.G. Ravenhall, \Journal{\PRC}
{58} {1804}{1998}.
\bibitem{leje} A. Lejeune, U. Lombardo, and W. Zuo, \Journal{\PL}{B
477}{45}{2000}.
\bibitem{zuo} W. Zuo, A. Lejeune, U. Lombardo, and J.F. Mathiot, preprint,
nucl-th/0202076. 
\bibitem{malf} B. ter Haar and R. Malfliet, \Journal{Phys. Rep.}{149}{207}
{1987}.
\bibitem{brock} R. Brockmann and R. Machleidt, \Journal{\PRC}
{42} {1965}{1990}.
\bibitem{fuchs} T. Gross-Boelting, C. Fuchs, and A. Faessler, \Journal{\NPA}
{648} {105}{1999}.
\bibitem{schil} E. Schiller and H. M\"uther, \Journal{Eur. Phys. J.}{A 11}{15}
{2001}.
\bibitem{fuji} J. Fujita and H. Miyazawa, \Journal{Prog. Theor. Phys.} {17}{360}
{1957}.
\bibitem{delta} M.R. Anastasio, A. Faessler, H. M\"uther, K. Holinde,  and R.
Machleidt, \Journal{\PRC}{18}{2416}{1978}.
\bibitem{shim} K. Shimizu, A. Polls, H. M\"uther, and A. Faessler, 
\Journal{\NPA}{364} {461}{1981}.
\bibitem{piper} S.C. Pieper, V. R. Pandharipande, R.B. Wiringa, and J. Carlson,
\Journal{\PRC}{64}{014001}{2001}.
\bibitem{report} H. M\"uther and A. Polls, \Journal{Prog. Part. and Nucl. Phys.}
{45}{243}{2000}.
\bibitem{heisen} J.H. Heisenberg and B. Mihaila, \Journal{\PRC}{59}{1440}{1999}.
\bibitem{idaho} D.R.\ Entem and R.\ Machleidt, \Journal{\PL}{B 524}{93}{2002}.
\bibitem{bbp} H.A.\ Bethe, B.H.\ Brandow, and A.G.\ Petschek, \Journal{\PREV}
{129}{225}{1963}.
\bibitem{kuem} H.~K\"ummel, K.~H.~L\"uhrmann, and J.~G.~Zabolitzky,
\Journal{\PREP} {36} {1} {1978}.
\bibitem{jeuk} J.P.\ Jeukenne, A.\ Lejeune, and C.\ Mahaux, \Journal{\PREP}{25}
{83} {1976}.
\bibitem{day} B.D. Day, \Journal{\PRC}{24}{1203}{1981}.
\bibitem{baldo} H.Q. Song, M. Baldo, G. Giansiracusa, and U. Lombardo,
\Journal{\PL}{B 411}{237}{1997};
\Journal{\PRL}{81}{1584}{1998};\\
M. Baldo, G. Gianrisacusa, U. Lombardo, and H.Q. Song, \Journal{\PL}{B 473}
{1}{2000}.
\bibitem{fiasco} M. Baldo and A. Fiasconaro, \Journal{\PL}{B 491}{240}{2000}.
\bibitem{khaga1} T. Frick, Kh. Gad, H. M\"uther, and P. Czerski, 
\Journal{\PRC}{65}{034321}{2002}. 
\bibitem{kuo1} T.T.S. Kuo, Z.Y. Ma, and R. Vinh Mau, \Journal{\PRC}{33}{717}
{1987}.
\bibitem{kuo2} M.F. Jing, T.T.S. Kuo, and H. M\"uther, \Journal{\PRC}{38}{2408}
{1988}.
\bibitem{skouras} H. M\"uther and L.D. Skouras, \Journal{\NPA}{555} {541}{1993}.
\bibitem{nili} K. Amir-Azimi-Nili, J.M. Udias, H. M\"uther, L.D. Skouras, and A.
Polls, \Journal{\NPA}{625} {633}{1997}.
\bibitem{spinorbit} L. Zamick, D.C. Zheng, and H. M\"uther, \Journal{\PRC}{45}
{2763}{1992}.
\bibitem{carlob} C. Barbieri and W.H. Dickhoff, preprint nucl-th/0111058.
\bibitem{hvh} N.M. Hugenholtz and L. Van Hove \Journal{Pysica}{24}{363}{1958}.
\bibitem{alfredo} P. Czerski, A. De Pace, and A. Molinari, \Journal{\PRC}{65}
{044317}{2002}. 
\bibitem{leusch} M. Leuschner, J.R. Calarco, F.W. Hersman, E. Jans, G.J. Kramer,
L. Lapikas, G. van der Steenhoven, P.K.A. de Witt Huberts, H.P. Blok, N.
Kalantar-Nayestanaki, and J. Friedrich, \Journal{\PRC}{49}
{955}{1994}. 
\end{thebibliography}
\end{document}